# A Soft Coordination Method of Heterogeneous Devices in Distribution System Voltage Control

Licheng Wang, Tao Wang, Gang Huang, *Member, IEEE*, Ruifeng Yan, *Member, IEEE*, Kai Wang, Youbing Zhang, *Member, IEEE*, and Shijie Cheng, *Life Fellow, IEEE*

*Abstract*—With the continuous increase of photovoltaic (PV) penetration, the voltage control interactions between newly installed PV inverters and previously deployed on-load tap-changer (OLTC) transformers become ever more significant. To achieve coordinated voltage regulation, current methods often rely on a decision-making algorithm to fully take over the control of all devices, requiring OLTC to give up its existing tap switching logic and execute corresponding upgrades. Aiming at bridging this gap, a soft coordination framework is proposed in this paper. Specifically, the decision-making commands are only applied on inverters, and OLTC that retains its own operation rule will be indirectly controlled by the changed system voltage, which is a result of appropriately adjusting inverters' Var output. The proposed method achieves the soft coordination by establishing a modified actor-critic algorithm to train a proxy model of inverters. The well-trained proxy model can properly adjust inverters' Var output to "softly" coordinate OLTC's tap operations, which finally attains coordinated voltage regulation and line loss minimization. Simulation results verify the superiority of our proposed method over traditional ones in coordinating heterogeneous devices for voltage control.

## I. INTRODUCTION

THE ever-rising photovoltaic (PV) penetration in distribution systems brings opportunities for carbon emission reduction as well as challenges of voltage regulation. Nowadays, voltage violation issues have become the main obstacle to further increase PV power integration in distribution systems.

In traditional distribution systems, system voltage regulation typically relies on tap switches of on-load tap-changer (OLTC) transformers deployed upstream. One of the most popular OLTC tap control logics is called line drop compensation (LDC). Specifically, an analogue circuit is used in OLTC to model the voltage drop of the distribution line. Therefore, OLTC can perceive remote voltage variations based on local measurements and accordingly adjusts its tap position if the estimated voltage is out of the allowable range for longer than the time delay [1, 2]. Other OLTC tap control logics with or without remote monitoring are reported in [3, 4]. These widely deployed OLTC transformers are originally designed to compensate for voltage changes caused by slow load variations, and they may remain effective if the PV penetration is not high [4, 5]. However, as the PV penetration continuously increases, OLTC transformers alone may not be able to successfully address the overvoltage issue any more due to substantial reverse power flow [6] and uneven PV power distribution [7]. Furthermore, fast PV power fluctuations may further induce excessive tap operations of OLTC transformers [8], which consequently accelerates the aging of devices.

Compared with OLTC transformers that typically have slow response speeds and discrete tap operations, inverters are capable of smoothly adjusting their Var output in real time, providing a more flexible way for voltage control. Efforts have been made to design coordinated voltage control methods for distributed PV inverters. For example, a distributed algorithm robust to communication asynchrony was proposed in [9] to optimize PV inverters' Var output for real time voltage control. The Volt-Var interaction across phases was investigated in [10], and an inter-phase coordinated voltage control method was designed for unbalanced distribution systems. Coordinated voltage control executed by inverter clusters is also included in but not limited to [11-13].

In most situations, PV inverters are integrated into distribution systems that have already been equipped with OLTC transformers for voltage regulation. Existing research for coordinated voltage control with heterogeneous devices can be mainly divided into two categories:

1) Rule-based coordination. Various coordination rules have been designed for heterogeneous devices to achieve coordinated voltage control. For example, in [14], the allowable voltage range was properly divided into several zones, and corrective actions from inverters or OLTC would be adaptively taken according to each zone. In [15], PV inverters and a battery energy storage system (BESS) were designed to be temporarily involved in voltage correction. They reduced their contribution in voltage control once an OLTC tap operation was triggered. Consequently, both PV inverters and the BESS were free from excessive usage. A

This work was supported by the National Natural Science Foundation of China (52007170, 52007173, U22B20116) and Zhejiang Provincial Natural Science Foundation of China (LY22E070007).

Licheng Wang, Tao Wang, and Youbing Zhang are with the College of Information Engineering, Zhejiang University of Technology, Hangzhou 310023, China (email: wanglicheng@zjut.edu.cn).

Gang Huang is with Zhejiang Lab, Hangzhou 311121, China, and also with Zhejiang University, Hangzhou 310027, China (corresponding author, e-mail: huanggang@zju.edu.cn).

Ruifeng Yan is with the School of Information Technology and Electrical Engineering, The University of Queensland, Brisbane, QLD 4072, Australia (e-mail: ruifeng@itee.uq.edu.au).

Kai Wang is with the College of Electrical Engineering, Qingdao University, Qingdao 266071, China (e-mail: wkwj888@163.com).

Shijie Cheng is with the School of Electrical and Electronic Engineering, Huazhong University of Science and Technology, Wuhan 430074, China (e-mail: sjcheng@hust.edu.cn).

coordination strategy to control BESS along with OLTC was proposed in [16]. The weighted average of estimated voltage at all buses was taken as the control signal to trigger OLTC tap operations, which led to balanced utilization of OLTC and BESS in voltage regulation. Similar research can also be seen in [17, 18].

2) Algorithm-based coordination. Both optimization and reinforcement learning (RL) based algorithms have been established to coordinate heterogeneous devices in voltage regulation. For example, a distribution system voltage control problem was cast into an RL framework in [19], where inverters, capacitor banks (CBs), and OLTC were cooperatively dispatched in the same time scale by deep Q-network (DQN). Considering diverse response speeds of different devices, a multi-timescale co-optimization model was formulated as a mixed-integer second-order cone program in [20], where network reconfiguration, OLTC, and inverters (battery and PV) were scheduled on daily, hourly, and 20-min bases respectively. Similarly in [21], OLTC, CBs, PVs, and mobile energy storage systems (MESS) were coordinated by a two-stage cost-effective control strategy for voltage regulation as well as cost minimization. Readers can refer to [22, 23] for more coordination schemes with multi-layer structures.

Despite their effectiveness in voltage regulation, rule-based coordination strategies often need empirical design. It is a more general way to coordinate heterogeneous devices by optimization or RL based algorithms. However, most current algorithm-based methods require full control of both newly installed inverters and previously deployed voltage regulation devices (e.g., OLTC transformers). As a result, these traditional devices, which have already been effectively running for many years, have to completely overturn their existing operation rules in order to be controlled by a new algorithm. Simultaneously, corresponding upgrades of traditional devices will make field implementation more complicated as well as bring extra costs.

To bridge this gap, a learning-based soft coordination control method that fully respects existing operation rules of previously deployed devices is proposed in this paper, with contributions summarized as follows:

1) Soft coordination framework: An innovative control framework that aims to "softly" collaborate inverters with OLTC transformers for system voltage regulation is proposed in Section II. Different from most current algorithm-based methods that rely on the direct control of all devices, in our soft coordination framework, OLTC, which operates with its existing control logic, will spontaneously participate in the coordinated voltage control with inverters. OLTC's tap operations can be indirectly controlled by inverters' Var output through coupled system voltage. In this paper, inverters will be controlled by a proxy model to delicately manage the system voltage through Var output to coordinate OLTC in a soft way.

2) Memory-based Markov decision process: Following the soft coordination framework, the coordinated voltage regulation problem is cast to a memory-based Markov decision process (MDP) as demonstrated in Section III. Specifically, proxy model-controlled inverters will properly output reactive power to interact with the OLTC-equipped distribution system (environment) and get corresponding rewards. Considering the time series-coupled tap switching mechanism of OLTC, the proxy model makes decisions for inverters according to the historical trajectory it has observed instead of current system states as in most traditional MDPs.

3) Recurrent soft actor-critic algorithm: With episodes obtained through inverter-environment interactions, the proxy model can be effectively trained by the recurrent soft actor-critic (RSAC) algorithm proposed in Section IV. The proxy model (called the actor network in the RSAC) is established as a Gated Recurrent Unit (GRU)-equipped deep neural network (DNN), which is designed to effectively process time series-coupled information and make proper decisions according to the historical trajectory. The well-trained proxy model (actor network) can properly adjust inverters' Var output to coordinate OLTC tap operations for successful voltage regulation and line loss minimization.

## II. Soft Coordination Framework

### A. Tap Control Logic of OLTC Transformers

OLTC transformers play a dominating role in voltage regulation in most distribution systems. Following its own operation rule, an OLTC transformer can adaptively adjust its tap position to compensate system voltage variations. Fig. 1 demonstrates one of the most popular tap control logics applied on OLTC transformers in the industry.

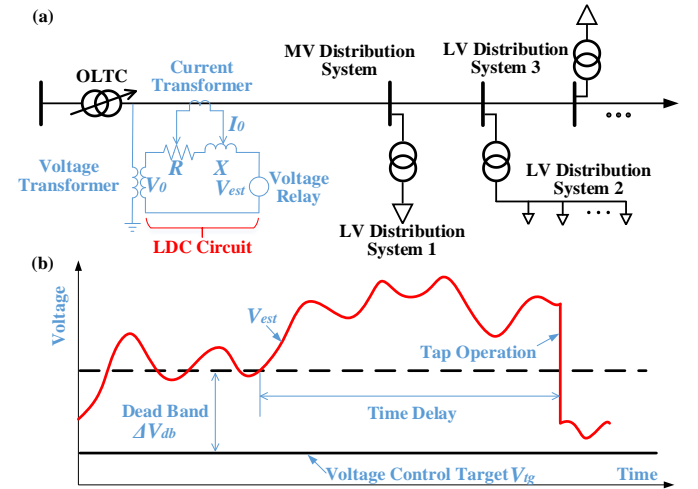

Fig. 1. A typical distribution system with an OLTC transformer for voltage regulation. (a) LDC circuit; (b) Tap operation.

As shown in Fig. 1 (a), an internal model called LDC circuit is used to match the distribution line impedance. Distribution system operators (DSOs) can set $R$ and $X$ values in the LDC circuit through the load-center method or voltage-spread method [1] to adjust the compensation. With the LDC circuit, a downstream voltage level $V_{est}$ can be estimated according to detected voltage $V_0$ and current $I_0$ on the secondary side of OLTC as:

$$V_{est} = V_0 - I_0(R + jX), \quad (1)$$

This estimated $V_{est}$ will be compared with the voltage control target $V_{tg}$ and dead band $\Delta V_{db}$. Once $V_{est}$ is out of its allowable range $[V_{tg} - \Delta V_{db}, V_{tg} + \Delta V_{db}]$, the timer of the OLTC transformer starts to count. As shown in Fig. 1 (b), temporary voltage violations will not trigger tap switches, since the OLTC timer will be interrupted once $V_{est}$ comes back to its allowable range. While, OLTC will step down (up) its tap position $Z_t$ if $V_{est}$ is larger (lower) than $V_{tg} + \Delta V_{db}$ ($V_{tg} - \Delta V_{db}$) and simultaneously the timer reading $TD_t$ exceeds the setting of time delay $T_d$. This time series-coupled LDC tap control logic is summarized as in Fig. 2, which has been widely used in OLTC transformers for adaptive tap switching.

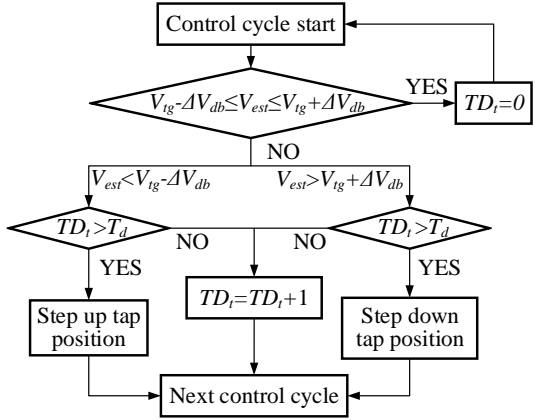

Fig. 2 The LDC based tap control logic of OLTC transformers.

### B. Interaction of Heterogeneous Devices in Voltage Control

Due to coupled system voltage, the voltage correction from inverters' Var output can be perceived by the upstream OLTC transformer, which inevitably leads to interactions between inverters and OLTC in system voltage regulation. As inverters can immediately adjust their Var output and thus change the system voltage, LDC equipped OLTC in a sense is under indirect control of inverters. For example, to avoid a tap switch, inverters will enhance its voltage correction so as to interrupt the OLTC timer before it arrives at the time delay $T_d$; Conversely, to trigger a tap switch, inverters need to weaken their voltage correction and allow a certain level voltage violation risks last longer than OLTC's time delay $T_d$.

Fig. 3 demonstrates distinct changes of OLTC behaviours with different strategies of inverters' Var output. If inverters are designed to keep their point of common coupling (PCC) voltage constant through Var compensation, no OLTC tap switch will be triggered, as in Fig. 3 (a). As a result, PV power induced overvoltage will be suppressed only by inverters' Var compensation for a long time, which consequently makes inverters vulnerable to Var saturation. Furthermore, significant reactive power flow will bring extra system line loss. If inverters relax their control of system voltage, for example, Volt-Var droop curves are applied on inverters, the PCC voltage will fluctuate with PV power variations. In this situation, OLTC tap operations will be triggered sometimes as in Fig. 3 (b), which in turn changes the operation points of inverters as well as mitigate their Var compensation burden. If inverters do not participate in voltage regulation at all, the OLTC transformer originally designed to compensate for slow changes of load has to directly face the fast-fluctuating PV power, which will thus cause excessive OLTC tap operations as in Fig. 3 (c) and make a transformer easier to be damaged.

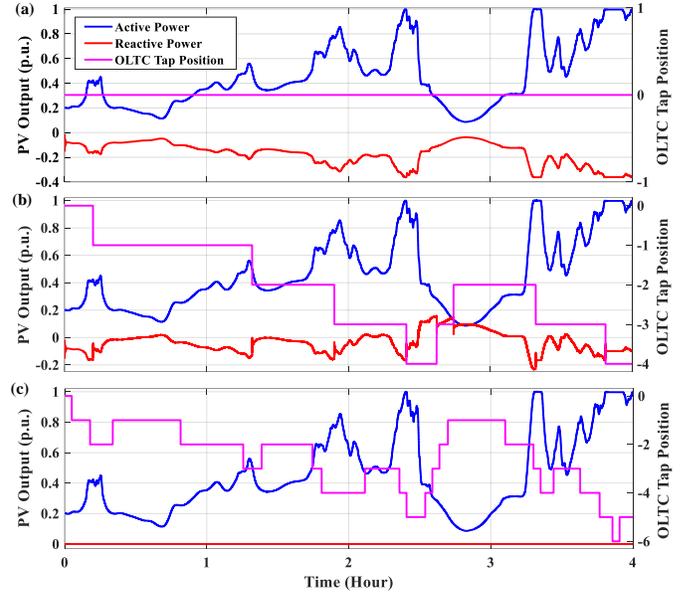

Fig. 3. Interactions of inverters and OLTC in system voltage regulation.

### C. Realization of Soft Coordination in Voltage Control

According to the discussion in Section II-B, OLTC has the potential to be indirectly controlled by inverters' Var output without overturning its current operation rule. Inspired by this idea, we propose an innovative soft coordination framework in this paper. Within this framework, the OLTC transformer will retain its existing control logic for tap switching, and proxy model-controlled inverters are able to "softly" collaborate OLTC's tap operations by appropriately adjusting their Var output, aiming to achieve coordinated system voltage regulation as well as line loss minimization.

To realize such a soft coordination mechanism, we formulate the voltage control problem as a memory-based MDP, where the proxy model-controlled inverters interact with the OLTC equipped distribution system (environment) through their Var output, as shown in Fig. 4. Instead of using current system states, the GRU-DNN based proxy model makes actions (Var output) for inverters according to the historical trajectory (i.e., a series of system states and actions) it has observed so far. As inverters adjust their Var output, the system states change, and the proxy model will receive a reward that evaluates its action value at the current time step. The changed system states $S_{t+1}$ and the corresponding action $A_t$ will then be fed back to the proxy model at time instant $t$ as a fragment of its observed historical trajectory. These time series-coupled inputs will be processed by GRU through recursive computation, which finally helps the DNN to make a proper decision in the next time step, as shown in Fig. 4.

The memory-based MDP as discussed above will be comprehensively introduced in Section III. With episodes obtained from inverter-environment interactions at different time steps, the GRU-DNN based proxy model can be effectively trained by our proposed RSAC algorithm to improve its performance in decision making. The GRU-DNN based proxy model as well as the established RSAC algorithm will be demonstrated in Section IV.

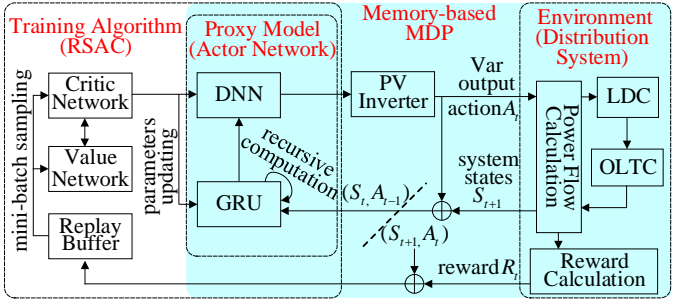

Fig. 4 The schematic figure of our proposed method.

## III. MEMORY-BASED MDP IN SYSTEM VOLTAGE CONTROL

### A. Tuple of Memory-based MDP in System Voltage Control

The voltage control problem within the soft coordination framework can be depicted by a memory-based MDP with the tuple $(\mathcal{A}, \mathcal{S}, \mathcal{R}, \mathcal{P})$, where $\mathcal{A}, \mathcal{S}, \mathcal{R}$, and $\mathcal{P}$ represent the sets of all possible actions, Markov states, rewards, and the probability distribution of state transition, respectively.

1) Actions

Within our designed soft coordination framework, the Var outputs of inverters are decision variables (actions). Therefore, at time instant $t$, the action $A_t \in \mathcal{A}$ is the set of the Var output of PV inverters $Q_t^{pv}$, namely $A_t = Q_t^{pv}$, and its upper and lower limits are $Q_{max}^{pv}$ and $-Q_{max}^{pv}$, respectively.

2) Markov States

According to the OLTC operation mechanism introduced in Section II-A, the voltage of all buses $V_t$, tap position $Z_t$, and the timer reading $TD_t$ (which indicates how long the over/under voltage risk has lasted) are necessary Markov states in depicting the OLTC behavior in its voltage control process. In addition, the system voltage is also influenced by active and reactive power flows. Therefore, the complete Markov states $S_t \in \mathcal{S}$ of the distribution system voltage control are given as:

$$S_t = (P_t^{load}, Q_t^{load}, P_t^{pv}, V_t, Z_t, TD_t), \quad (2)$$

where $P_t^{load}$, $Q_t^{load}$, $P_t^{pv}$ represent the sets of active load, reactive load, and active power output of PV inverters, respectively, at time instant $t$. It is worth noting, $P_t^{load}$, $Q_t^{load}$, $P_t^{pv}$ are uncontrollable in the voltage control problem, while $Q_t^{pv}$ is the set of decision variables (actions).

3) Rewards

Our designed soft coordination has two control aims: a) voltage violation elimination, b) line loss minimization. Correspondingly, if there are voltage violations at any buses, the reward function is designed as (3) to quantify the system voltage deviation. Otherwise, the reward function is as (4) to minimize system line loss.

$$R_t = M \sum_{j \in \mathcal{N}} [max(V_t^j - \overline{V}, 0) + max(\underline{V} - V_t^j, 0)] \quad (3)$$

$$R_t = \lambda(P_t^{loss,0} - P_t^{loss}), \quad (4)$$

where $R_t \in \mathcal{R}$ is the reward; $M < 0$ represents the penalty coefficient of voltage violations; $\overline{V}$ and $\underline{V}$ are upper and lower limits of system voltage, respectively; $V_t^j$ denotes the voltage of bus $j$ at time instant $t$; $\mathcal{N}$ is the set of all buses; $\lambda > 0$ represents the incentive factor; $P_t^{loss}$ represents the line loss if action $A_t$ is taken at time instant $t$, and $P_t^{loss,0}$ corresponds to the one that no action is taken.

4) Probability Distribution of State Transitions

The system state will transit from $S_t$ to $S_{t+1}$ with a reward $R_t$ after an action $A_t$ is taken at time instant $t$. Such a state transition obeys a probability distribution $\rho \in \mathcal{P}$, denoted as below:

$$(S_{t+1}, R_t) \sim \rho(\cdot | S_t, A_t). \quad (5)$$

### B. Historical Trajectory and Decision Making

Current system states are not sufficient for making appropriate decisions to achieve soft coordination, because the operation mechanism of OLTC transformers is time series-coupled, as demonstrated in Section II-A. To address this issue, the historical trajectory $K_t \in \mathcal{K}$, defined in (6), is used as the input for the proxy model's decision making process.

$$K_t = (\underbrace{S_0, A_0, S_1, A_1, \ldots, S_{t-1}}_{K_{t-1}}, A_{t-1}, S_t), \quad (6)$$

where the historical trajectory $K_t$ represents a series of states and actions obtained from inverter-environment interactions from the initial moment to time instant $t$, and $K_t$ is comprised of $K_{t-1}$ and $(A_{t-1}, S_t)$. The proxy model uses the observed historical trajectory $K_t$ to output a probability distribution of actions $\pi_\phi(\cdot | K_t)$, instead of directly providing a specific action $A_t$. The action $A_t$ is then randomly selected based on the obtained probability distribution, as shown in (7):

$$A_t \sim \pi_\phi(\cdot | K_t), \quad (7)$$

where $\pi_\phi$ represents the policy function, and $\phi$ is the network parameter. The input-output relationship of the proxy model is established by the policy function, which maps the historical trajectory $K_t$ to the action probability distribution $\pi_\phi(\cdot | K_t)$.

### C. MDP Considering Entropy

Different from standard deep reinforcement learning (DRL) that aims only to maximize the expectation of accumulated rewards, we introduce an entropy term [24] into our learning objective (8). It is worth noting that larger entropy values correspond to more random actions. With maximum entropy learning, the range of possible actions can be sufficiently explored, and premature convergence to local optimum could be avoided.

The learning objective is defined as follows:

$$J = \mathop{\mathbb{E}}_{\tau \sim \pi_\phi} \left\{ \sum_{n=0}^{N} \gamma^n \left[ R_{t+n} + \alpha \mathcal{H}\left(\pi_\phi(\cdot | K_{t+n})\right) \right] \right\}, \quad (8)$$

where $\tau$ represents future trajectories following a policy $\pi_\phi$, $N$ denotes the length of trajectories, $\alpha > 0$ is a temperature parameter used to balance the rewards and entropy, and $\gamma \in (0,1)$ is the discount factor. Herein, $\mathcal{H}\left(\pi_\phi(\cdot|K_t)\right)$ denotes the entropy of the probability distribution of actions for a given policy $\pi_\phi$ and historical trajectory $K_t$.

According to the learning objective in (8), the future performance of a policy $\pi_\phi$ with a given historical trajectory $K_t = k_t$ can be depicted by the value function $V^{\pi_\phi}$, which is defined in (9):

$$V^{\pi_\phi}(k_t) = \underset{\tau \sim \pi_\phi}{\mathbb{E}} \left\{ \sum_{n=0}^{N} \gamma^n \left[R_{t+n} + \alpha \mathcal{H}\left(\pi_\phi(\cdot|K_{t+n})\right)\right] |_{K_t = k_t} \right\}. \tag{9}$$

Herein, $V^{\pi_\phi}(k_t)$ represents the expected value of the accumulated rewards and entropy in the future trajectory starting from $K_t = k_t$ if policy $\pi_\phi$ is applied for action making.

Similarly, the action value function $Q^{\pi_\phi}$ is defined in (10):

$$Q^{\pi_\phi}(k_t, a_t) = \underset{\tau \sim \pi_\phi}{\mathbb{E}} \left\{ \sum_{n=0}^{N} \gamma^n R_{t+n} + \alpha \sum_{n=1}^{N} \gamma^n \mathcal{H}\left(\pi_\phi(\cdot|k_{t+n})\right) |_{K_t = k_t, A_t = a_t} \right\}. \tag{10}$$

Herein, $Q^{\pi_\phi}(k_t, a_t)$ represents the expected value of the accumulated rewards and entropy that can be obtained by following policy $\pi_\phi$ after taking action $A_t = a_t$ based on a known historical trajectory $K_t = k_t$.

## IV. RECURRENT SOFT ACTOR-CRITIC ALGORITHM

### A. Structure of the RSAC Algorithm

The proxy model needs to be sufficiently trained using our proposed RSAC algorithm to maximize the accumulated rewards that it can achieve. The RSAC algorithm is based on the soft actor-critic (SAC) algorithm [24], but with an improved ability to process time series-coupled information. As the name suggests, the RSAC algorithm includes an actor network and a critic network. The actor network is responsible for making actions, and the critic network estimates the values of those actions. It is important to note that the proxy model being trained in this paper acts as the actor network in the RSAC algorithm. Therefore, the proxy model will be referred to as the action network henceforth. In this section, we will introduce the actor network, the critic network, and other auxiliaries of the RSAC algorithm.

1) Actor Network

As shown in Fig. 4, the actor network (proxy model) is established as a GRU-equipped DNN, which is designed to output proper actions (Var compensation) for inverters based on the historical trajectory $K_t$ (actor network's input). Herein, GRU is a type of recurrent neural networks (RNNs) with a gating mechanism [25]. Such refinement of RNN includes an update gate and a reset gate, which determine what information is allowed through to the output, and it can be trained to retain information over time. With this simulated ability to remember information, GRU is used in this paper to process the time-series coupled variable-length historical trajectory $K_t$ as:

$$H_t = g_{\phi_r}(S_t, A_{t-1}, H_{t-1}), \tag{11}$$

where $g_{\phi_r}$ denotes the GRU mapping rule, and $\phi_r$ represents its network parameter. $H_t$ represents the GRU hidden state at time instant $t$, which will be updated at each time step. Through recursive computations as presented in (11), a variable-length historical trajectory $K_t$ is projected into a fixed-dimension $H_t$. The updated $H_t$ is then transferred to a DNN as its input, and its output is the probability distribution of actions. In this paper, possible actions $A_t$ are designed to obey a normal distribution, and correspondingly the DNN has a $2N_A$-dimensional output as:

$$(\mu_t, \sigma_t) = f_{\phi_d}(H_t), \tag{12}$$

where $N_A$ is the dimension of action $A_t$; $\mu_t \in \mathbb{R}^{1 \times N_A}$ and $\sigma_t \in \mathbb{R}^{1 \times N_A}$ are the sets of expectation and variance, respectively; $\phi_d$ represents the DNN parameter; $f_{\phi_d}$ denotes the DNN mapping rule. So far, the policy $\pi_\phi$ of the actor network has been parameterized as:

$$(\mu_t, \sigma_t) = \pi_\phi(\widetilde{H}_t) = f_{\phi_d}[g_{\phi_r}(S_t, A_{t-1}, H_{t-1})], \tag{13}$$

where $\phi = [\phi_r, \phi_d]$; $(S_t, A_{t-1}, H_{t-1})$ is denoted by $\widetilde{H}_t$ for ease of description. According to the normal distribution given by (13), action $A_t$ can finally be obtained through random sampling as:

$$A_t = \mathbb{P}(\mu_t + \varepsilon_t \odot \sigma_t), \tag{14}$$

where $\varepsilon_t$ denotes the random variable obeying a standard normal distribution $\mathcal{N}(0,1)$, and $\odot$ is the dot product operator. Since the inverters' Var output (i.e., action $A_t$) has physical boundaries, $\mathbb{P}$ represents the projection from infinity to the interval $[-Q_{max}^{pv}, Q_{max}^{pv}]$.

2) Critic Network

In this paper, the action value function $Q^{\pi_\phi}$ defined in (10) is approximated by a neural network called critic network $Q_\theta^{\pi_\phi}$ (with parameter $\theta$). The critic network $Q_\theta^{\pi_\phi}$ as in (15) acts as a critic to evaluate the value of an action $A_t$ taken by the policy $\pi_\phi$ (actor network) based on a known $\widetilde{H}_t$. Specifically, with a given policy $\pi_\phi$, the critic network $Q_\theta^{\pi_\phi}$ can give out a Q value to criticize the action $A_t$ based on known $\widetilde{H}_t$.

$$Q = Q_\theta^{\pi_\phi}(\widetilde{H}_t, A_t). \tag{15}$$

3) Value Network

Similarly, the value function defined in (9) can be approximated by the value network $V_\psi^{\pi_\phi}$ with parameter $\psi$ as:

$$Q = V_\psi^{\pi_\phi}(\widetilde{H}_t). \tag{16}$$

With a given policy $\pi_\phi$, the value network $V_\psi^{\pi_\phi}$ can give out a Q value to estimate the expectation of reward and entropy that can be accumulated over the future based on the current $\widetilde{H}_t$.

4) Replay Buffer

The latest $N_b$ sets of the actor network's experiences (called episodes), which are a series of tuples $(\widetilde{H}_t, A_t, R_t, S_{t+1})$ obtained from inverter-environment interactions, are stored in the replay buffer $\mathcal{D}$. A mini-batch of episodes will be randomly sampled from the replay buffer and used for each round of parameter updating.

## B. Network Parameter Updating

In our proposed RSAC algorithm, the parameters of the actor network ($\phi$) and the critic network ($\theta$) are alternately and iteratively updated by the strategies of Policy Evaluation and Policy Improvement, respectively, while the value network works as an auxiliary in parameter updating. Through Policy Evaluation, the critic network $Q_\theta^{\pi_\phi}$ with updated $\theta$ could have better performance in action value estimation for a given policy $\pi_\phi$. The improved critic network $Q_\theta^{\pi_\phi}$ will further be used to update the actor network ($\phi$) through Policy Improvement, and the updated actor network (policy $\pi_{\phi'}$) will have better performance in action selection. Namely, actions made by updated policy $\pi_{\phi'}$ tend to have larger Q values compared with that of the original policy $\pi_\phi$.

### 1) Critic Network Updating Through Policy Evaluation

The value function $V^{\pi_\phi}$ and action value function $Q^{\pi_\phi}$ as defined in (9) and (10), respectively, yield the following Bellman equations:

$$Q_\theta^{\pi_\phi}(\widetilde{H}_t, A_t)$$
$$= \mathop{\mathbb{E}}_{\tau \sim \pi_\phi} \left\{ \sum_{n=0}^{N} \gamma^n R_{t+n} + \alpha \sum_{n=1}^{N} \gamma^n \mathcal{H}\left(\pi_\phi(\cdot|\widetilde{H}_{t+n})\right) \right\}$$
$$= \mathop{\mathbb{E}}_{\tau \sim \pi_\phi} \left\{ R_t + \sum_{n=1}^{N} \gamma^n [R_{t+n} + \alpha \mathcal{H}\left(\pi_\phi(\cdot|\widetilde{H}_{t+n})\right)] \right\}$$
$$= \mathop{\mathbb{E}}_{\tau \sim \pi_\phi} \left\{ R_t + \gamma \sum_{n=1}^{N} \gamma^{n-1} [R_{t+n} + \alpha \mathcal{H}\left(\pi_\phi(\cdot|\widetilde{H}_{t+n})\right)] \right\}$$
$$= \mathop{\mathbb{E}}_{(S_{t+1}, R_t) \sim \rho} \left\{ R_t + \gamma V_\psi^{\pi_\phi}(\widetilde{H}_{t+1}) \right\} \quad (17)$$

$$V_\psi^{\pi_\phi}(\widetilde{H}_t) = \mathop{\mathbb{E}}_{\tau \sim \pi_\phi} \left\{ \sum_{n=0}^{N} \gamma^n \left[ R_{t+n} + \alpha \mathcal{H}\left(\pi_\phi(\cdot|\widetilde{H}_{t+n})\right) \right] \right\}$$
$$= \mathop{\mathbb{E}}_{\tau \sim \pi_\phi} \left\{ \sum_{n=0}^{N} \gamma^n R_{t+n} + \alpha \sum_{n=1}^{N} \gamma^n \mathcal{H}\left(\pi_\phi(\cdot|\widetilde{H}_{t+n})\right) + \alpha \mathcal{H}\left(\pi_\phi(\cdot|\widetilde{H}_t)\right) \right\}$$
$$= \mathop{\mathbb{E}}_{A_t \sim \pi_\phi} \left\{ Q_\theta^{\pi_\phi}(\widetilde{H}_t, A_t) - \alpha \log P_{\pi_\phi}(A_t|\widetilde{H}_t) \right\}. \quad (18)$$

It is worth noting that the variable-length historical trajectory $K_t$ (as in value function $V^{\pi_\phi}$ and action value function $Q^{\pi_\phi}$) has been projected into a fixed-dimension $H_t$ in the value network $V_\psi^{\pi_\phi}$ and critic network $Q_\theta^{\pi_\phi}$, where $\widetilde{H}_t = (S_t, A_{t-1}, H_{t-1})$ for ease of description. In addition, the entropy $\mathcal{H}\left(\pi_\phi(\cdot|\widetilde{H}_t)\right)$ in this paper is defined as:

$$\mathcal{H}\left(\pi_\phi(\cdot|\widetilde{H}_t)\right) = -\mathop{\mathbb{E}}_{A_t \sim \pi_\phi} \left\{ \log P_{\pi_\phi}(A_t|\widetilde{H}_t) \right\}, \quad (19)$$

where $P_{\pi_\phi}(A_t|\widetilde{H}_t)$ represents the probability density of action $A_t$ in the probability distribution $\pi_\phi(\cdot|\widetilde{H}_t)$.

Equations (17) and (18) are used to derive two loss functions as follows:

$$J_Q(\theta) = \mathop{\mathbb{E}}_{(\widetilde{H}_t, A_t, R_t, S_{t+1}) \sim \mathcal{D}} \left\{ \left[ Q_\theta^{\pi_\phi}(\widetilde{H}_t, A_t) - R_t - \gamma V_{\bar\psi}^{\pi_\phi}(\widetilde{H}_{t+1}) \right]^2 \right\} \quad (20)$$

$$J_V(\psi) = \mathop{\mathbb{E}}_{\widetilde{H}_t \sim \mathcal{D}, A_t \sim \pi_\phi} \left\{ \left[ V_\psi^{\pi_\phi}(\widetilde{H}_t) - Q_\theta^{\pi_\phi}(\widetilde{H}_t, A_t) + \alpha \log P_r(A_t|\widetilde{H}_t) \right]^2 \right\}, \quad (21)$$

where $(\widetilde{H}_t, A_t, R_t, S_{t+1}) \sim \mathcal{D}$ means that the tuple $(\widetilde{H}_t, A_t, R_t, S_{t+1})$ is randomly sampled from the replay buffer $\mathcal{D}$, and $\widetilde{H}_t \sim \mathcal{D}$ means that $\widetilde{H}_t$ is randomly sampled from $\mathcal{D}$. It is worth noting that the target value network $V_{\bar\psi}^{\pi_\phi}$ rather than the value network $V_\psi^{\pi_\phi}$ is used in (20) to improve the stability of the algorithm.

With the derived loss functions in (20) and (21), the parameters $\theta$ and $\psi$ can be iteratively updated through mini-batch stochastic gradient decent:

$$\theta \leftarrow \theta + \eta \nabla_\theta J_Q(\theta) \quad (22)$$
$$\psi \leftarrow \psi + \eta \nabla_\psi J_V(\psi). \quad (23)$$

During the updating of $\theta$, the parameter $\bar\psi$ in (20) is kept constant, and it is periodically synchronized with $\psi$ as:

$$\bar\psi \leftarrow \beta \bar\psi + (1 - \beta)\psi. \quad (24)$$

Herein, parameters $\eta, \beta \in (0,1)$ in (22)~(24) represent the learning rates used in the mini-batch stochastic gradient descent optimization algorithm.

### 2) Actor Network Updating Through Policy Improvement

After sufficient training, the critic network $Q_\theta^{\pi_\phi}$ can output $Q$ values for action value estimation, where a larger $Q$ value indicates a better action. To improve the performance of the actor network, one direct idea is to have a probability distribution of actions, such that actions with larger $Q$ values correspond to larger probability densities, and vice versa. Therefore, the target probability distribution $\pi_{\phi'}(\cdot|\widetilde{H}_t)$ given by the actor network is aimed to be shaped as:

$$\pi_{\phi'}(\cdot|\widetilde{H}_t) \rightarrow \frac{\exp\{Q_\theta^{\pi_\phi}(\widetilde{H}_t, \cdot)/\alpha\}}{\int \exp\{Q_\theta^{\pi_\phi}(\widetilde{H}_t, A_t)/\alpha\} dA_t}, \quad (25)$$

where exp{} represents the exponential operation, ensuring that $\exp\{Q_\theta^{\pi_\phi}(\widetilde{H}_t, \cdot)/\alpha\}$ is always nonnegative (since probability density is always nonnegative), and $Q_\theta^{\pi_\phi}(\widetilde{H}_t, A_t) > Q_\theta^{\pi_\phi}(\widetilde{H}_t, A'_t)$ always leads to $\exp\{Q_\theta^{\pi_\phi}(\widetilde{H}_t, A_t)/\alpha\} > \exp\{Q_\theta^{\pi_\phi}(\widetilde{H}_t, A'_t)/\alpha\}$. In addition, with such a design, $\int \pi_{\phi'}(A_t|\widetilde{H}_t) dA_t = 1$ always holds, which is consistent with the physical significance that the integral of probability density of all possible actions is always equal to 1.

The Kullback-Leibler (KL) divergence is used to measure the deviation between two distributions. Correspondingly, the training objective is designed to minimize the KL divergence as:

$$\phi' \leftarrow \arg\min_{\phi'} D_{KL} \left\{ \pi_{\phi'}(\cdot|\widetilde{H}_t) \middle\| \frac{\exp\{Q_\theta^{\pi_\phi}(\widetilde{H}_t, \cdot)/\alpha\}}{\int \exp\{Q_\theta^{\pi_\phi}(\widetilde{H}_t, A_t)/\alpha\} dA_t} \right\}, \quad (26)$$

where $D_{KL}$ represents the KL divergence between two distributions. Since the value of $\int exp\{Q_\theta^{\pi_\phi}(\widetilde{H}_t, A_t)/\alpha\} dA_t$ in (26) is independent of the selection of action $A_t$, it can be denoted as $F_\theta^{\pi_\phi}(\widetilde{H}_t)$ for ease of description. According to (26), the loss function is given as:

$$J_\pi(\phi') = D_{KL}\left\{\pi_{\phi'}(\cdot|\widetilde{H}_t) \left\| \frac{exp\{Q_\theta^{\pi_\phi}(\widetilde{H}_t,\cdot)/\alpha\}}{F_\theta^{\pi_\phi}(\widetilde{H}_t)} \right.\right\}$$
$$= \int P_{\pi_{\phi'}}(A_t|\widetilde{H}_t) \log \frac{P_{\pi_{\phi'}}(A_t|\widetilde{H}_t)}{exp\{Q_\theta^{\pi_\phi}(\widetilde{H}_t,A_t)/\alpha\}/F_\theta^{\pi_\phi}(\widetilde{H}_t)} dA_t$$
$$= \mathop{\mathbb{E}}_{\widetilde{H}_t \sim \mathcal{D}, A_t \sim \pi_{\phi'}} \left\{\log P_{\pi_{\phi'}}(A_t|\widetilde{H}_t) + \log F_\theta^{\pi_\phi}(\widetilde{H}_t) - \frac{Q_\theta^{\pi_\phi}(\widetilde{H}_t,A_t)}{\alpha}\right\}. \quad (27)$$

By substituting (14) into (27), we can equivalently express the loss function as:

$$J_\pi(\phi') = \mathop{\mathbb{E}}_{\widetilde{H}_t \sim \mathcal{D}, \varepsilon \sim \mathcal{N}(0,1)} \left\{\log P_{\pi_{\phi'}}(\mathbb{P}(\mu_t + \varepsilon_t \odot \sigma_t)|\widetilde{H}_t) + \log F_\theta^{\pi_\phi}(\widetilde{H}_t) - Q_\theta^{\pi_\phi}(\widetilde{H}_t, \mathbb{P}(\mu_t + \varepsilon_t \odot \sigma_t))/\alpha\right\}. \quad (28)$$

Finally, the parameter $\phi$ of the action network can be updated through the mini-batch stochastic gradient decent method:

$$\phi \leftarrow \phi + \eta \nabla_\phi J_\pi(\phi). \quad (29)$$

## V. CASE STUDIES

### A. Experiment Settings

A modified IEEE 33-bus distribution system with a 16-step OLTC transformer and 7 distributed PV inverters, as shown in Fig. 5, is used for case studies. As introduced in Section II-A, the OLTC transformer follows its LDC rule for system voltage regulation with control parameters given in Table I. The maximum active and reactive power of PV inverters on different buses are listed in Table II. The allowable range of the system voltage is set to be 0.95p.u.~1.05p.u. in this paper.

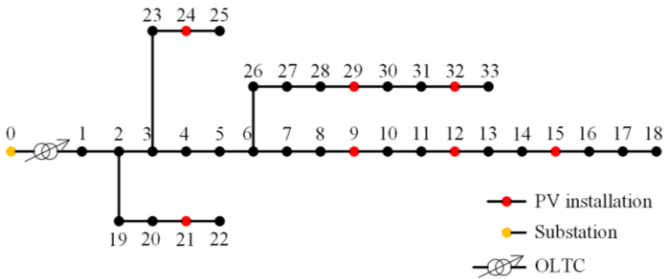

Fig. 5. Topology of the modified IEEE 33-bus system.

TABLE I
CONTROL PARAMETERS OF THE OLTC TRANSFORMER

| $V_{tg}$ | $R$ | $X$ | $\Delta V_{db}$ | $T_d$ | regulator range |
|---|---|---|---|---|---|
| 1 (p.u.) | 0.864 (p.u.) | 0.538 (p.u.) | 0.008 (p.u.) | 180s | $\pm 5\%$ |

TABLE II
MAXIMUM ACTIVE AND REACTIVE POWER OF PV INVERTERS

| Bus | 9 | 12 | 15 | 21 | 24 | 29 | 32 |
|---|---|---|---|---|---|---|---|
| $P_{max}$ | 600kW | 600kW | 1000kW | 400kW | 400kW | 600kW | 1000kW |
| $Q_{max}$ | 240kVar | 240kVar | 400kVar | 160kVar | 160kVar | 240kVar | 400kVar |

The 33-bus distribution system model is programmed in OpenDSS, and the proposed RSAC algorithm is implemented in Python using PyTorch [26] on a 64-bit machine with 3.70GHz CPU and 16GB RAM. The Component Object Model (COM) is used for the information exchange between OpenDSS and Python. The hyperparameters of the RSAC algorithm are shown in Table III.

TABLE III
HYPERPARAMETERS OF THE PROPOSED ALGORITHM

| Parameter | Value |
|---|---|
| optimizer | Adam |
| network type | feed-forward & recurrent |
| non-linearity | ReLU |
| size of hidden layers | {256, 256} |
| mini batch size | 256 |
| replay buffer size | $10^5$ |
| $\eta$ | $3*10^{-3}$ |
| $\beta$ | $10^{-2}$ |
| $\gamma$ | 0.95 |
| $M$ | $-10^2$ |
| $\lambda$ | $10^2$ |

### B. Training Process

Fig. 6 shows the episode average reward value during the training process of successive 1500 episodes. Initially, in the early learning phase, the action policies lead to negative rewards due to limited positive learning experiences and unoptimized action policies. These negative rewards illustrate that the actor network is incapable of maintaining the system voltages within 0.95p.u.~1.05p.u. and simultaneously reducing system line loss. However, as the training progresses, the actor network gradually evolves and obtains positive rewards more frequently. A positive reward implies that there is no voltage violation, and the system line loss is further reduced by taking actions. It is observed that the episode average reward keeps fluctuating, but with an upward trend. The training process converges after about 1000 episodes.

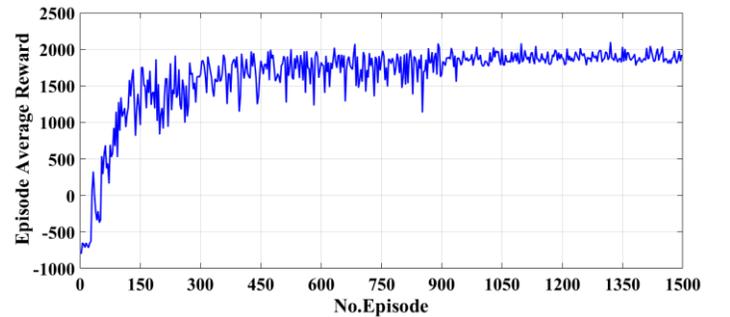

Fig. 6. Training process of the RSAC algorithm.

## C. Baseline Methods

To evaluate the proposed method's voltage regulation performance and system line loss, two representative methods are implemented as baselines. The details are as follows:

1) *Baseline-1*: A two-stage method is used to coordinate the upstream OLTC transformer and downstream PV inverters, aiming to minimize the system line loss and control the system voltage within the allowable range. In the first stage, OLTC tap positions for every 15-minute period are scheduled in advance by solving an optimization problem based on the day-long forecast of PV power. On this basis, operation points of PV inverters are optimized by a soft actor-critic (SAC) algorithm for real time Volt-Var control in the second stage. However, it is worth noting that the previously installed OLTC transformer has to fully overturn its existing tap switching rule and be upgraded accordingly to be dispatched by this two-stage method.

2) *Baseline-2*: The widely used Volt-Var droop curves are applied to PV inverters for local voltage control. With such a scheme, the previously deployed OLTC transformer can retain its existing rule for voltage regulation. Consequently, device upgrade costs can be saved, and this method is easy for field implementation. However, neither line loss reduction nor coordinated operation between the OLTC transformer and PV inverters is considered.

## D. Strong PV Power Fluctuating Scenario

In the Baseline-1, the 15-minute average of the ground truth PV power profile is regarded as the prediction for day-long OLTC tap operation scheduling, as shown in Fig. 7. Based on its optimal power flow (OPF) model, Baseline-1 aims to minimize the power loss by increasing the system voltage level. Consequently, OLTC tap positions of Baseline-1 are much higher than that of Baseline-2 and our proposed method, as shown in Fig. 8. However, due to fast moving cloud coverage on cloudy days, PV power may experience significant changes within 15 minutes, as shown in Fig. 7. As a result, the online voltage control of inverters in the second stage, based on OLTC tap positions scheduled in the first stage, may not always be able to completely mitigate overvoltage issues, as shown in Fig. 9. Except for Baseline-1, both Baseline-2 and our proposed method can successfully control the system voltage during days with strong PV power fluctuations.

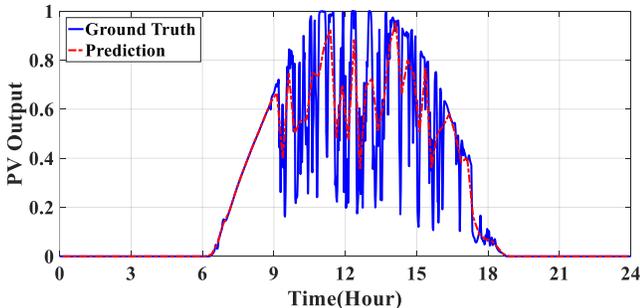
Fig. 7. PV power profile with strong fluctuations and its 15-minute prediction.

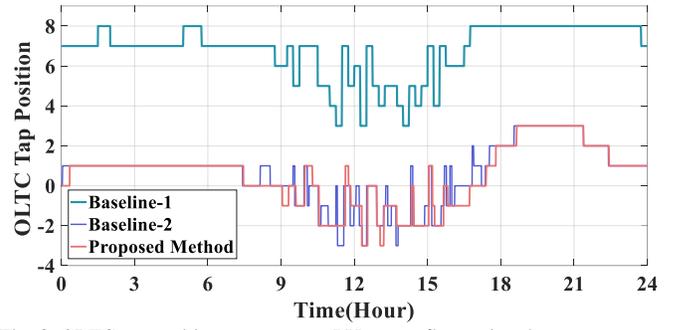
Fig. 8. OLTC tap positions on a strong PV power fluctuating day.

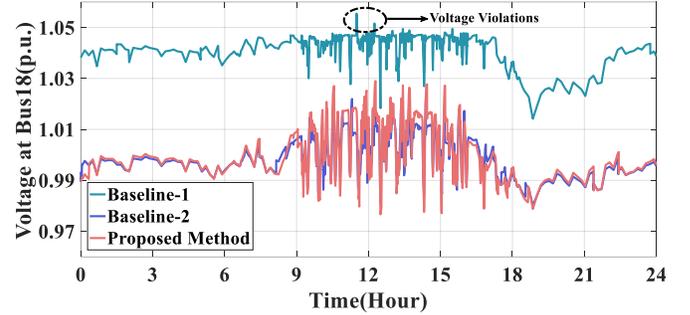
Fig. 9. Voltage profiles at bus 18 on a strong PV power fluctuating day.

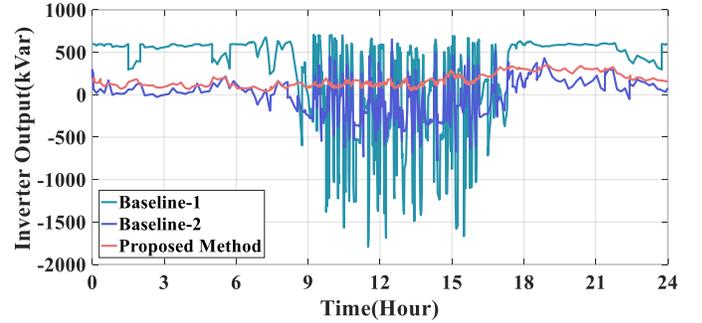
Fig. 10. Inverters' Var compensation on a strong PV power fluctuating day.

As designed in Baseline-1, inverters' Var output is used as the second stage control to compensate for the control mismatch of the first stage's OLTC scheduling in voltage regulation. Since Baseline-1 schedules OLTC with higher tap positions, inverters have to absorb a large amount of reactive power to suppress the corresponding overvoltage if the real time PV power is larger than its every 15-minute prediction. As shown in Fig. 10, Baseline-1 needs much more Var compensation compared with Baseline-2 and our proposed method during days the strong PV power fluctuations. In addition, extra reactive power flow will increase system line loss, resulting in Baseline-1 having the largest power loss (625.82kWh) during a 24-hour period, as shown in Table IV. In contrast, for Baseline-2, inverters' Var output fluctuates with the system voltage according to the adopted droop control curve. In our proposed method, the Var output of inverters is appropriately adjusted by the well-trained proxy model to achieve soft coordination with OLTC tap operations, aiming to maximize the accumulated reward in the long term. Consequently, our proposed method can successfully control the system voltage with the least power loss (559.39kWh) among all three methods. Finally, Table V compares the number of OLTC tap operations of different methods during the day. Our proposed method triggers 30 tap operations, while there will be 31 and 47 tap operations, respectively, if Baseline-1 and Baseline-2 are applied on the same day.

TABLE IV
SYSTEM LINE LOSS OF DIFFERENT METHODS

| Baseline-1 | Baseline-2 | Proposed Method |
|---|---|---|
| 625.82kWh | 584.86kWh | 559.39kWh |

TABLE V
NUMBERS OF OLTC TAP OPERATIONS OF DIFFERENT METHODS

| Baseline-1 | Baseline-2 | Proposed Method |
|---|---|---|
| 31 | 47 | 30 |

### E. Mild PV Power Fluctuating Scenario

Fig. 11 demonstrates a day-long PV power profile with only mild fluctuations. Compared with strong PV power fluctuating days, the 15-minute average fits the ground truth PV power profile better in this scenario. Consequently, if the 15-minute average PV power is regarded as the prediction for OLTC tap position scheduling in the first stage, the two-stage method (i.e., Baseline-1) is expected to perform better.

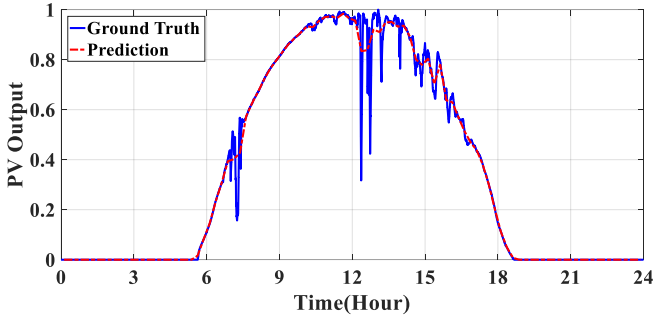
Fig. 11. PV power profile with mild fluctuations and its 15-minute prediction.

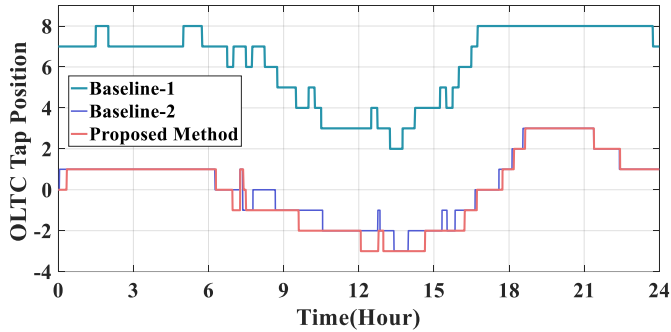
Fig. 12. OLTC tap positions on a mild PV power fluctuating day.

The day-long OLTC tap positions of different methods are compared in Fig. 12. Similar to the situation in the strong PV power fluctuating scenario, the OPF-based Baseline-1 has the highest OLTC tap positions during the day, which leads to the highest voltage profile shown in Fig. 13. Thanks to the relatively accurate prediction as well as mild PV power fluctuations, inverters' online Var compensation in the second stage of Baseline-1 can successfully control the system voltage in this scenario. Compared with Baseline-1, both Baseline-2 and our proposed method have lower OLTC tap positions and voltage profiles during the same day. Fig. 14 compares inverters' Var compensation profiles of different methods, and our proposed method needs the least Var compensation for system voltage control on the mild PV power fluctuation day.

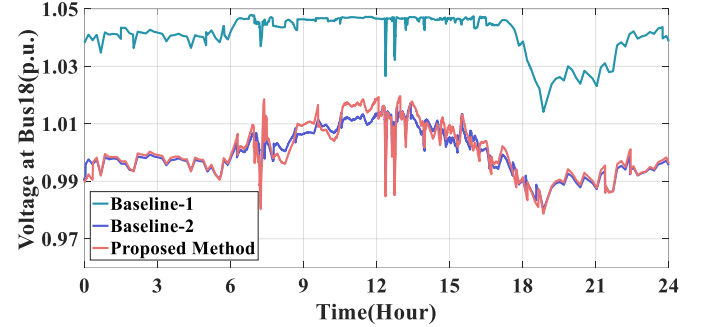
Fig.13. Voltage profiles at bus 18 on a mild PV power fluctuating day.

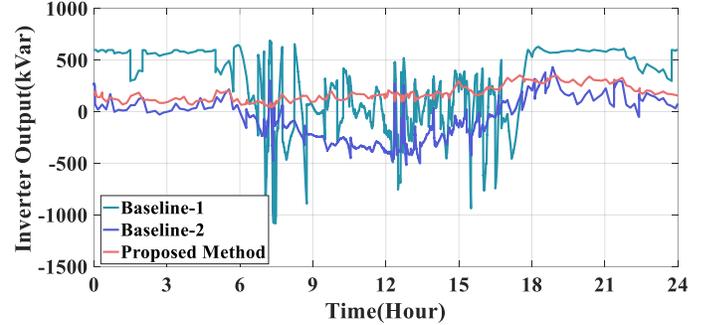
Fig.14. Inverters' Var compensation on a mild PV power fluctuating day.

The system line loss and numbers of OLTC tap operations of all three methods are compared in Table VI and Table VII, respectively. As PV power fluctuations become mild, the gaps in system line loss of different methods are reduced compared to the strong PV power fluctuating scenario. As shown in Table VI, our proposed method has the minimum line loss (732.27kWh), and the maximum line loss is Baseline-2's 758.40kWh in this scenario. Simultaneously, the numbers of OLTC tap operations of Baseline-2 and our proposed method are significantly reduced from 47 to 20 and from 30 to 18, respectively. It is worth noting that the OLTC transformer follows its LDC control rule in both Baseline-2 and our proposed method, and it can adaptively increase or reduce its tap operations if the PV power fluctuations become strong or mild. However, the OLTC tap positions in Baseline-1 are scheduled by its OPF method based on a 15-minute PV power prediction, which cannot fully reflect the real PV power fluctuations during a day. As a result, the number of OLTC tap operations of Baseline-1 does not have a significant change in strong and mild PV power fluctuating scenarios.

TABLE VI
SYSTEM LINE LOSS OF DIFFERENT METHODS

| Baseline-1 | Baseline-2 | Proposed Method |
|---|---|---|
| 743.35kWh | 758.40kWh | 732.27kWh |

TABLE VII
NUMBERS OF OLTC TAP OPERATIONS OF DIFFERENT METHODS

| Baseline-1 | Baseline-2 | Proposed Method |
|---|---|---|
| 26 | 20 | 18 |

## VI. CONCLUSION

This paper proposes an innovative control framework for coordinating inverters and OLTC to regulate system voltage in a "soft" manner. Within this control framework, previously deployed OLTC can join the coordinated voltage regulation without requiring any device upgrades, and system line loss minimization can be achieved simply by adjusting the Var output of inverters. Therefore, compared with most optimal coordination methods that need to take control of all involved devices, our proposed method is more cost-effective and easier to implement.

The voltage regulation performance and system line loss of our proposed method are compared with two baseline methods in case studies. Baseline-1 is a two-stage method whose performance depends on the accuracy of PV power prediction, as future PV power profiles are needed for optimal OLTC tap scheduling. In Baseline-2, inverters are controlled by local Volt-Var droop curves, and OLTC follows its existing tap switching control logic. Although easy to apply, inverters and OLTC do not operate in a coordinated manner in Baseline-2. Our proposed method has the advantage of being easy to implement, as it does not require forecast information or OLTC upgrades. Simulation results indicate that our proposed method outperforms the baseline methods in terms of voltage regulation, system line loss, and the number of OLTC tap operations in both strong and mild PV power fluctuation scenarios.

## REFERENCES

[1] T. A. Short, *Electric Power Distribution Equipment and Systems*, United States of America: CRC Press, 2006.
[2] W. H. Kersting, *Distribution System Modeling and Analysis*: CRC Press, 2002.
[3] C. Long, and L. F. Ochoa, "Voltage Control of PV-Rich LV Networks: OLTC-Fitted Transformer and Capacitor Banks," *IEEE Transactions on Power Systems,* vol. 31, no. 5, pp. 4016-4025, 2016.
[4] A. T. Procopiou, and L. F. Ochoa, "Voltage Control in PV-Rich LV Networks Without Remote Monitoring," *IEEE Transactions on Power Systems,* vol. 32, no. 2, pp. 1224-1236, 2017.
[5] N. Yorino, Y. Zoka, M. Watanabe, and T. Kurushima, "An Optimal Autonomous Decentralized Control Method for Voltage Control Devices by Using a Multi-Agent System," *IEEE Transactions on Power Systems,* vol. 30, no. 5, pp. 2225-2233, 2015.
[6] M. I. Hossain, R. Yan, and T. K. Saha, "Investigation of the interaction between step voltage regulators and large-scale photovoltaic systems regarding voltage regulation and unbalance," *IET Renewable Power Generation,* vol. 10, no. 3, pp. 299-309, 2016.
[7] L. Wang, T. K. Saha, and R. F. Yan, "Voltage Regulation for Distribution Systems with Uneven PV Integration in Different Feeders," *2017 IEEE Power & Energy Society General Meeting*, 2017.
[8] L. Wang, R. F. Yan, and T. K. Saha, "Voltage Management for Large Scale PV Integration into Weak Distribution Systems," *IEEE Transactions on Smart Grid,* vol. 9, no. 5, pp. 4128-4139, Sep, 2018.
[9] J. Li, C. Liu, M. E. Khodayar, M.-H. Wang, Z. Xu, B. Zhou, and C. Li, "Distributed Online VAR Control for Unbalanced Distribution Networks With Photovoltaic Generation," *IEEE Transactions on Smart Grid,* vol. 11, no. 6, pp. 4760-4772, 2020.
[10] L. Wang, R. Yan, F. Bai, T. K. Saha, and K. Wang, "A Distributed Inter-Phase Coordination Algorithm for Voltage Control with Unbalanced PV Integration in LV Systems," *IEEE Transactions on Sustainable Energy,* vol. 11, no. 4, pp. 2687-2697, Oct, 2020.
[11] H. Liu, and W. Wu, "Two-Stage Deep Reinforcement Learning for Inverter-Based Volt-VAR Control in Active Distribution Networks," *IEEE Transactions on Smart Grid,* vol. 12, no. 3, pp. 2037-2047, 2021.
[12] L. Wang, L. Xie, Y. Yang, Y. Zhang, K. Wang, and S.-j. Cheng, "Distributed Online Voltage Control with Fast PV Power Fluctuations and Imperfect Communication," *IEEE Transactions on Smart Grid*, early access, DOI: 10.1109/TSG.2023.3236724, 2023.
[13] Y. Wang, M. H. Syed, E. Guillo-Sansano, Y. Xu, and G. M. Burt, "Inverter-Based Voltage Control of Distribution Networks: A Three-Level Coordinated Method and Power Hardware-in-the-Loop Validation," *IEEE Transactions on Sustainable Energy,* vol. 11, no. 4, pp. 2380-2391, 2020.
[14] M. r. Jafari, M. Parniani, and M. H. Ravanji, "Decentralized Control of OLTC & PV Inverters for Voltage Regulation in Radial Distribution Networks with High PV Penetration," *IEEE Transactions on Power Delivery*, 2022.
[15] L. Wang, F. Bai, R. Yan, and K. T. Saha, "Real-Time Coordinated Voltage Control of PV Inverters and Energy Storage for Weak Networks With High PV Penetration," *IEEE Transactions on Power Systems,* vol. 33, no. 3, pp. 3383-3395, May, 2018.
[16] T. Tewari, A. Mohapatra, and S. Anand, "Coordinated Control of OLTC and Energy Storage for Voltage Regulation in Distribution Network With High PV Penetration," *IEEE Transactions on Sustainable Energy,* vol. 12, no. 1, pp. 262-272, 2021.
[17] M. Chamana, B. H. Chowdhury, and F. Jahanbakhsh, "Distributed Control of Voltage Regulating Devices in the Presence of High PV Penetration to Mitigate Ramp-Rate Issues," *IEEE Transactions on Smart Grid,* vol. 9, no. 2, pp. 1086-1095, 2018.
[18] K. M. Muttaqi, A. D. T. Le, M. Negnevitsky, and G. Ledwich, "A Coordinated Voltage Control Approach for Coordination of OLTC, Voltage Regulator, and DG to Regulate Voltage in a Distribution Feeder," *IEEE Transactions on Industry Applications,* vol. 51, no. 2, pp. 1239-1248, 2015.
[19] Y. Zhang, X. Wang, J. Wang, and Y. Zhang, "Deep Reinforcement Learning Based Volt-VAR Optimization in Smart Distribution Systems," *IEEE Transactions on Smart Grid,* vol. 12, no. 1, pp. 361-371, 2021.
[20] R. Zafar, and H. R. Pota, "Multi-Timescale Coordinated Control with Optimal Network Reconfiguration using Battery Storage System in Smart Distribution Grids," *IEEE Transactions on Sustainable Energy*, pp. 1-12, 2023.
[21] X. Sun, J. Qiu, Y. Yi, and Y. Tao, "Cost-Effective Coordinated Voltage Control in Active Distribution Networks With Photovoltaics and Mobile Energy Storage Systems," *IEEE Transactions on Sustainable Energy,* vol. 13, no. 1, pp. 501-513, 2022.
[22] Q. Yang, G. Wang, A. Sadeghi, G. B. Giannakis, and J. Sun, "Two-Timescale Voltage Control in Distribution Grids Using Deep Reinforcement Learning," *IEEE Transactions on Smart Grid,* vol. 11, no. 3, pp. 2313-2323, 2020.
[23] Y. Huo, P. Li, H. Ji, H. Yu, J. Yan, J. Wu, and C. Wang, "Data-Driven Coordinated Voltage Control Method of Distribution Networks With High DG Penetration," *IEEE Transactions on Power Systems,* vol. 38, no. 2, pp. 1543-1557, 2023.
[24] T. Haarnoja, A. Zhou, P. Abbeel, and S. Levine. "Soft Actor-Critic: Off-Policy Maximum Entropy Deep Reinforcement Learning with a Stochastic Actor," https://arxiv.org/pdf/1801.01290.pdf.
[25] K. Cho, B. van Merrienboer, D. Bahdanau, and Y. Bengio. "On the Properties of Neural Machine Translation: Encoder-Decoder Approaches."
[26] PyTorch, available: https://pytorch.org/.